\documentclass[prd,a4paper,nofootinbib,showpacs,twocolumn
]{revtex4}
\usepackage{graphicx}
\usepackage{amsfonts}
\usepackage{color}

\def\hbar{\hspace{0pt}\raisebox{1pt}{$-$} \hspace{-7pt} h}

\def\5{\overline 5}

\newcommand{\be}{\begin{equation}}
\newcommand{\ee}{\end{equation}}
\newcommand{\bea}{\begin{eqnarray}}
\newcommand{\eea}{\end{eqnarray}}

\newcommand{\ba}{\begin{eqnarray}}
\newcommand{\ea}{\end{eqnarray}}

\begin{document}
\title[]{The electroweak $S$ and $T$ parameters from a fixed point condition}

\author{M. Fabbrichesi$^{\ddag}$}
\author{R. Percacci$^{\dag\ddag}$}
\author{A. Tonero$^{\dag\ddag}$}
\author{L. Vecchi$^{*}$}
\affiliation{$^{\ddag}$INFN, Sezione di Trieste} 
\affiliation{$^{\dag}$SISSA,
via Bonomea 265, 34136 Trieste, Italy}
\affiliation{$^{*}$LANL, Los Alamos, NM 87545, USA}

\begin{abstract}
We consider the standard model without the Higgs boson,
where the Goldstone modes are described by a nonlinear sigma model.
We study the renormalization group flow of the sigma model coupling $\tilde f$
and of the electroweak parameters $S$ and $T$. 
The condition that the couplings reach a fixed point
at high energy leaves the low energy values of $\tilde f$ and $T$ arbitrary 
(to be determined experimentally) and
fixes $S$ to a value compatible with electroweak precision data.
\end{abstract}
\maketitle
%

The nonlinear sigma model with values in a coset space $G/H$ arises whenever
a symmetry $G$ is spontaneously broken to $H$.
The best known application is the chiral model with $G=SU(2)\times SU(2)$
and $H=SU(2)$ (the diagonal, or vector subgroup), which describes the
low energy dynamics of pions, regarded as the (pseudo)-Goldstone bosons
arising from spontaneous breaking of chiral symmetry in the theory of strong interactions with two massless fermion flavors.
An equally important realization of the same geometry describes the
Goldstone bosons that break the electroweak (EW) $SU(2)\times U(1)$ to $U(1)$.
In this case the Goldstone bosons do not correspond to physical states,
rather they are transformed into the longitudinal components of the $W$ and $Z$
bosons by the Higgs phenomenon. Essentially all we currently know about
EW interactions can be encoded in 
an effective field theory of Goldstone bosons coupled to gauge fields and fermions
\cite{ab}.
This would be the minimal option: in fact it would have no Higgs boson
in the Lagrangian and hence one {\it less} degree of freedom than the 
standard model (SM).

Due to its perturbative nonrenormalizability, the nonlinear sigma model 
is usually regarded as a mere low energy effective field theory.
In fact, in the case of strong interactions, the UV completion of the
chiral model is QCD, so there is no reason to look further.
In the EW case, however, things are not yet settled and it is important
to consider all options.
The simplest possibility is to embed the nonlinear sigma model into
a complex doublet transforming linearly under $SU(2)_L$;
this renders the theory perturbatively renormalizable 
(though not UV complete, due to the positive beta function of the scalar coupling).
Technicolor provides a dynamical way of breaking the EW group.
Another possibility that we shall consider here 
is that the theory is renormalizable in a nonperturbative sense,
namely at a nontrivial fixed point (FP) of the renormalization group (RG) flow \cite{Weinberg}.
This idea has been developed mostly in the context of gravity \cite{reviews}.
For other applications to the SM see \cite{Gies}.
This approach has the disadvantage that perturbation theory is at best a rough guide.
If in spite of this one considers the one loop beta functions,
or some resummation thereof, it is easy to see that a nontrivial FP is present
\cite{codello}.
It persists when one considers in addition terms with four derivatives
of the Goldstone bosons \cite{hasenfratz} or the coupling to gauge fields \cite{fptz}.
In the same spirit, we will consider here
the compatibility of this hypothesis with precision EW data.
The effect of physics beyond the SM on the gauge bosons can be tested
by calculating the oblique parameters $S$ and $T$ \cite{peskin} 
and comparing with their experimental bounds.
We will study this issue by calculating the RG flow of the effective couplings
representing these parameters in the EW effective theory.

We will restrict ourselves to the bosonic sector of the EW effective theory.
Fermions may change the picture significantly; they will be considered
in a forthcoming paper \cite{fptvz}.
We use a geometrical description of the Goldstone bosons as coordinates
$\varphi^\alpha(x)$ of a field $U(x)$ taking values in
$SU(2)\times U(1)/U(1)\sim SU(2)$.
The lowest order terms in the Euclidean action are 
\bea
\label{action}
S_{\rm } &=& 
\frac{1}{2f^2} \int d^4x\, h_{\alpha\beta}D_\mu\varphi^\alpha D^\mu\varphi^\beta
\nonumber\\
&&
+ \frac{1}{4g^2}\int d^4x\, W^I_{\mu\nu}W_I^{\mu\nu} + \frac{1}{4g^{\prime 2}}\int d^4x\, B_{\mu\nu}B^{\mu\nu}
\eea
where $W^I_\mu$ are the $SU(2)_L$ gauge fields and $B_\mu$ is the $U(1)_Y$ gauge field;
$g$ and $g^\prime$ are the gauge couplings and 
$f$ is the (dimension $-1$) chiral coupling.
The covariant derivative acting on $\varphi$ is 
\be
D_\mu\varphi^\alpha = \partial_\mu\varphi^\alpha + W^I_\mu R^\alpha_I-B_\mu L^\alpha_3\, ,
\ee
while the gauge field strength tensors are 
$W^I_{\mu\nu}=\partial_\mu W_\nu^I-\partial_\nu W_\mu^I+\epsilon^I_{JM}W_\mu^JW_\nu^M$ and $B_{\mu\nu}=\partial_\mu B_\nu-\partial_\nu B_\mu$.
The indices $\alpha,\beta = 1,2,3$ run over the target space coordinates while $I,J,M = 1,2,3$ are $SU(2)$ Lie-algebra indices.
We denote $R^\alpha_I$ and $L^\alpha_I$ the right- and left-invariant vectorfields
on $SU(2)$. In particular, $R^\alpha_I$ generate $SU(2)_L$ and $L^\alpha_3$
generates $U(1)_Y$.

The gauge invariance of the SM demands that the metric $h_{\alpha\beta}$
be invariant under the action of these vectorfields, but not necessarily
under the $SU(2)_R$ transformations generated by $L^\alpha_1$ and $L^\alpha_2$.
The most general metric of this type is of the form
\be
h_{\alpha\beta}=L_\alpha^1 L_\beta^1+L_\alpha^2 L_\beta^2+(1-2a_0) L_\alpha^3 L_\beta^3\ ,
\ee
where $L_\alpha^I$ is the basis of left-invariant one-forms dual to $L^\alpha_I$.
The parameter $a_0$ measures the violation
of the ``custodial'' symmetry $SU(2)_R$ and vanishes in the bare SM Lagrangian.
Radiative corrections then induce a small nonvanishing effective value for $a_0$.
It is therefore customary to assume that the metric $h_{\alpha\beta}$ is bi-invariant
and to consider the $SU(2)_R$-breaking as due to a separate term in the 
effective Lagrangian:
\be
\label{Tterm}
\frac{a_0}{f^2} ({\rm tr}\sigma_3 U^\dagger DU)^2
=\frac{a_0}{f^2}  D_\mu\varphi^\alpha D^\mu\varphi^\beta L_\alpha^3 L_\beta^3\, .
\ee
The action contains further terms. Among these we shall be interested
in particular in the term
\be
\label{Sterm}
a_1\frac{1}{2} B_{\mu\nu}{\rm tr}\sigma_3 U^\dagger W^{\mu\nu} U
=a_1\frac{1}{2} B^{\mu\nu} W_{\mu\nu}^I R_{I\alpha}L^\alpha_3\, .
\ee
These definitions agree with those of \cite{feruglio,herreroruiz},
except for the rescaling of the gauge fields with the gauge couplings.
The running couplings $a_0$ and $a_1$ are related to the oblique parameters $S$ and $T$ by
\bea
\label{sdef}
S&=&-16\pi a_1(m_Z)
+\frac{1}{6\pi}\left[\frac{5}{12}-\log\left(\frac{m_H}{m_Z}\right)\right]
\\
\label{tdef}
T&=&\frac{2}{\alpha}a_0(m_Z)
-\frac{3}{8\pi\cos^2\theta_W}\left[\frac{5}{12}-\log\left(\frac{m_H}{m_Z}\right)\right]\,.
\eea
The second term on the r.h.s. corresponds to subtracting the contribution
of the Higgs field with mass $m_H$ \cite{bagger}.

In this paper we will be concerned with the RG running of the gauge couplings $g$, $g'$,
the sigma model coupling $f$ and the parameters $a_0$ and $a_1$.

It will be instructive to consider first the ungauged $SU(2)\times U(1)/U(1)$
sigma model, with couplings $f$ and $a_0$.
Quite generally, the beta function of the sigma model is given by a kind of Ricci flow
\cite{codello}:
\be
\label{zetadot}
\frac{d}{dt}\left(\frac{1}{f^2} h_{\alpha\beta}\right)
=\frac{1}{(4\pi)^{2}}k^2 R_{\alpha\beta}\, ,
\ee
where $t=\log k$.
In the basis of the right-invariant vectorfields,
the Ricci tensor of the metric $h_{\alpha\beta}$ is
$R_{11}=R_{22}=\frac{1}{2}+a_0$, $R_{33}=\frac{1}{2}-a_0$,
so the beta functions of $\tilde f^2=f^2 k^2$ and $a_0$ are
\bea
\label{betaf}
\frac{d\tilde f^2}{dt}&=&2\tilde f^2-\frac{1}{(4\pi)^2}\tilde f^4\left(\frac{1}{2}+a_0\right)
\\
\label{betaa0}
\frac{da_0}{dt}&=&\frac{1}{2}\frac{1}{(4\pi)^2}\tilde f^2 a_0(1-2a_0)\ .
\eea
These beta functions admit a Gaussian FP with $\tilde f=0$
and arbitrary $a_0$, and two nontrivial fixed points:
an $SU(2)_R$-symmetric one at $a_0=0$, $\tilde f=8\pi\approx 25.13$
and another one with strongly broken $SU(2)_R$ at $a_0=1/2$, $\tilde f=4\sqrt{2}\pi\approx 17.8$.
The FP at $a_0=0$ is UV-repulsive, the one at $a_0=1/2$ is UV-attractive.
If $a_0<0$, corresponding to an elongated three-sphere, 
$a_0$ decreases with increasing energy;
If $0<a_0<1/2$, corresponding to a mildly squashed three-sphere, 
$a_0$ increases with energy towards the FP at $a_0=1/2$;
If $a_0>1/2$, corresponding to a strongly squashed three-sphere, 
$a_0$ decreases with energy towards the FP at $a_0=1/2$.

Coming to the gauged case, we begin by considering the subsystem 
of the couplings $g$, $g'$ and $f$, keeping $a_0=a_1=0$.
This is a slight generalization of a calculation described in detail in \cite{fptz}.
The beta functions of the gauge couplings are
\bea
\label{gdot}
\frac{dg^2}{dt}&=&
\frac{g^4}{(4\pi)^2}
\frac{1}{1+\tilde m_W^2}
\left[-\frac{16}{(1+\tilde m_W^2)^2}
+\frac{3}{2}
\right]
\\
\label{gprimedot}
\frac{dg^{\prime 2}}{dt}&=&\frac{1}{6}
\frac{g^{\prime 4}}{(4\pi)^2}\frac{1}{1+\tilde m_W^2}
\eea
where $\tilde m_W^2=m_W^2/k^2=g^2/\tilde f^2$. The fractions represent the effect of thresholds and automatically switch off the beta functions when $k$ becomes smaller than $m_W$.
Aside from these thresholds, the difference with the SM
is due only to the absence of the Higgs particle, and is quite small,
so $g$ is asymptotically free, while $g'$ has a Landau pole at a trans-Planckian energy.

The beta function of $\tilde f^2$ is
\begin{widetext}
\bea
\frac{d\tilde f^2}{dt}&=&2\tilde f^2
-\frac{1}{(4\pi)^2}\Biggl\{
\frac{1}{4}\frac{\tilde f^4}{(1+\tilde m_W^2)^2}
+\frac{1}{4}\frac{\tilde f^4}{(1+\tilde m_Z^2)^2}
+\frac{2g^2\tilde f^2}{(1+\tilde m_W^2)^3}
\nonumber
\\
&&
+\frac{g^{\prime 2}\tilde f^2}{(1+\tilde m_W^2)(1+\tilde m_B^2)}
\left[\frac{1}{(1+\tilde m_W^2)}+\frac{1}{(1+\tilde m_B^2)}\right]
+\frac{g^2\tilde f^2}{(1+\tilde m_W^2)(1+\tilde m_Z^2)}
\left[\frac{1}{(1+\tilde m_W^2)}+\frac{1}{(1+\tilde m_Z^2)}\right]\Biggr\}
 \, ,
\eea%
\end{widetext}
where $\tilde m_Z^2=m_Z^2/k^2=(g^2+g^{\prime 2})/\tilde f^2$
and we also define the shorthand $\tilde m_B^2=g^{\prime 2}/\tilde f^2$.
The whole expression simplifies drastically when the mass terms can be neglected.
In practice one can use this approximation when $k>m_Z$, the heaviest mass in the theory,
while for $k<m_B=g'/f$, the lightest mass in the theory, the beta function reduces
to the first (classical) term.
In the following we will use this approximation.

Because of the positive beta function for $g'$, strictly speaking
this system does not have a FP.
However, the running of the gauge couplings is very slow
and for our purposes it is a good approximation to treat them as constants.
Setting $g=0.65$, $g'=0.35$ we find an approximate UV-attractive FP at $\tilde f=25.08$.
As expected it is very close to the FP of the ungauged model.

We are now ready to consider the effect of the couplings $a_0$ and $a_1$.
As in the ungauged case, the beta functions of $f$ and $a_0$ can be
extracted from the geometric beta functional of the metric.
For $k$ much larger than all the masses ($g,g'\ll\tilde f$), 
the threshold fractions become equal to one and the beta functions simplify to
\bea
\label{betaf2}
\frac{d\tilde f^2}{dt}&=&2\tilde f^2
-\frac{1}{2}\frac{\tilde f^2}{(4\pi)^2}\left(
\tilde f^2(1+2a_0)
+6g^2+3g^{\prime 2}\right)\,\, 
%
\\
\label{betaa02}
\frac{da_0}{dt}&=&
\frac{1}{2}\frac{1}{(4\pi)^{2}}
\left(
\tilde f^2a_0(1-2a_0)
+\frac{3}{2}g^{\prime 2}
\right)
\, .
\eea
We have neglected terms of order $g^2 a_0$ or $g^{\prime 2}a_0$,
which are subleading relative to those of order $\tilde f^2 a_0$.
They are not necessarily subleading relative to the terms of
order $g^2$ and $g^{\prime 2}$ that have been written, 
but they would be unimportant in what follows.
Note that these beta functions reduce correctly to 
(\ref{betaf}) and (\ref{betaa0}) in the ungauged case.
The first term in (\ref{betaa02}) corresponds to a self-renormalization of the operator (\ref{Tterm}).
Diagrammatically it corresponds to a quadratically divergent Goldstone boson tadpole
and cannot be seen in dimensional regularization.
The second term agrees with the results of \cite{herreroruiz};
it is proportional to $g^{\prime 2}$, consistent with the fact that the
hypercharge coupling breaks the custodial symmetry.
Its effect is to generate a nonzero $a_0$ even if initially $a_0=0$.

The fixed points of the ungauged case are slightly shifted by the
gauge couplings. They occur at: 
(FPI) $\tilde f=25.1$, $a_0=-0.000292$
and (FPII) $\tilde f=17.7$, $a_0=0.501$. 
There is no longer a fixed point with $\tilde f=0$.
This flow is illustrated in Fig.1.

\begin{figure}
{\resizebox{0.8\columnwidth}{!}
{\includegraphics{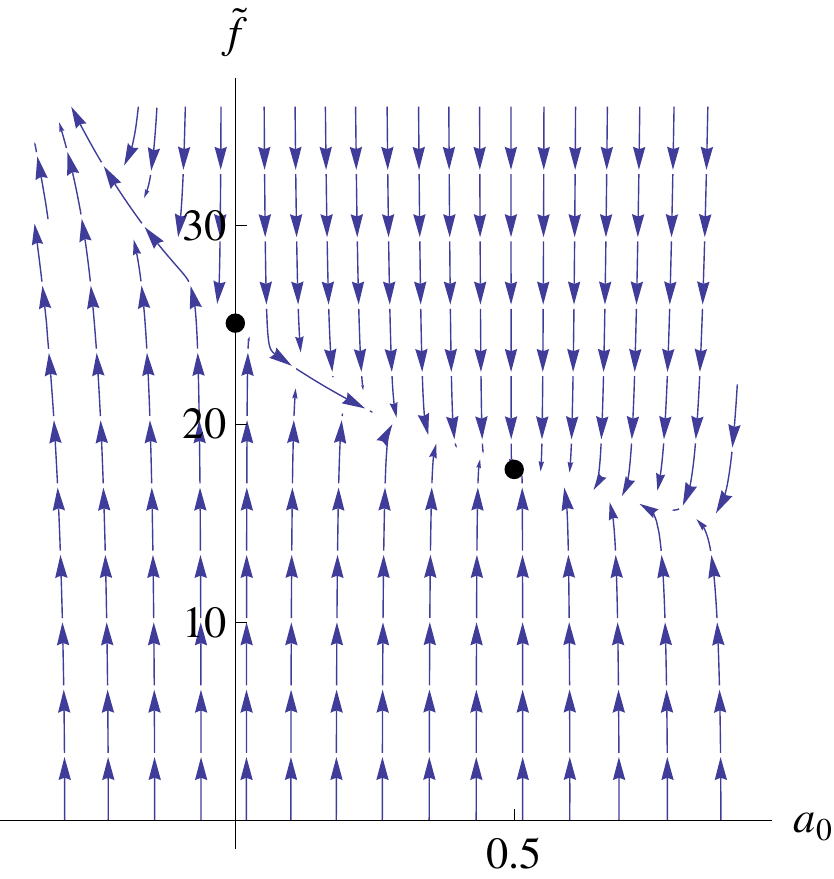}}
\caption{\small Flow in the $a_0$-$\tilde f$ plane. The two dots mark the positions of FPI and FPII.
Arrows point to increasing energy.
}
}
\end{figure}

The beta function of $a_1$ is
\be
\frac{da_1}{dt}=\frac{1}{(4\pi)^2}\left(\tilde f^2 a_1+\frac{1}{6}\right)\, .
\ee
Also in this case the second term agrees with the one computed in \cite{herreroruiz},
while the first comes from the self-renormalization of the operator (\ref{Sterm}).
Introducing the FP values for $\tilde f^2$ discussed above, we find
the FP values  $a_1=-0.000265$ for FPI and $a_1=-0.000530$ for FPII.
The eigenvalues and eigenvectors of the matrix describing the linearized
flow around these FP's are given by the following table:
\bigskip
\begin{center}
\begin{tabular}{|c|c|c|c|c|}
\hline
 FP             &eigenvalue & \multispan3 \hfil eigenvector components  \vline \\
 \hline
  &    & $\tilde f$ & $a_0$  & $a_1$  \\
\hline
I & $-1.99$ & $1.00$ & $11.6\times 10^{-6}$  &  $14.1\times 10^{-6}$  \\
\hline
I & $1.99$ & $-0.997$ & $0.0795$  &  $-42.2\times 10^{-6}$  \\
\hline
I & $3.98$ & $0$ & $0$  &  $1$  \\
\hline
II  & $-1.99$ & $1.00$ & $66.0\times 10^{-6}$  & $29.9\times 10^{-6}$  \\
\hline
II  & $-0.996$ & $-0.998$  & $0.0563$  & $-40\times 10^{-6}$  \\
\hline
II   & $1.99$ & $0$  & $0$  &  $1$ \\
\hline
\end{tabular}
\end{center}
\bigskip
Recall that negative eigenvalues correspond to UV attractive (relevant) directions.
The point FPI has one such direction, that to
a good approximation can be identified with the parameter $\tilde f$. 
The point FPII has two relevant directions that lie almost
exactly in the $a_0$-$\tilde f$ plane.
Within numerical errors we found a critical trajectory that starts
from FPII in the UV approximately in the direction of (minus) its second eigenvector and 
reaches FPI in the IR from the direction of its second eigenvector.
The origin is not a FP, but the beta functions become very small there.
This almost-FP is IR attractive for $\tilde f$.

We now discuss the physics of these FPs.
At $k=m_Z$ we have $\tilde f=2 m_Z/\upsilon=0.7415$ and
the experimentally allowed values for $a_0(m_Z)$ and $a_1(m_Z)$ are of order $10^{-3}$.
When one evolves the flow towards higher energies,
$\tilde f$, $a_0$ or $a_1$ will generally diverge.
This is a sign that ``new physics'' has to be taken into account.
However, there may be trajectories that hit a FP in the UV:
for them the effective field theory description actually never breaks down.
Such trajectories are said to be ``renormalizable'' or ``asymptotically safe'' 
\cite{Weinberg} and they form the so-called ``UV critical surface'',
which in the vicinity of a FP is spanned by the relevant couplings.

Requiring that the world be described by a renormalizable trajectory
leads to predictions for low energy physics.
Since FPI has only one relevant direction, there is a single renormalizable trajectory
that descends from it towards the origin.
Since the beta functions go to zero for $k<m_Z$, we stop the flow
at the scale $m_Z$ (i.e. when $\tilde f=0.7415$) and find, at that scale,
\be
\label{point}
a_0(m_Z)=-0.0020,\ \ a_1(m_Z)=-0.0032\ ,
\ee
which are 5$\sigma$ away from the experimental values.
The transition takes about four or five $e$-foldings (a change in scale by a factor $e^4$-$e^5$) 
which means that FPI would be reached at an energy scale of the order of 10 TeV.

The point FPII has two relevant directions and therefore there is a one parameter
family of renormalizable trajectories that descend from it.
From Fig.1 we see that for such a trajectory to come close to the origin,
it has to be fine tuned to first follow very closely the critical trajectory towards FPI,
and hence descend.
Going upwards from $k=m_Z$, such a trajectory would take again four or five $e$-foldings to reach
the vicinity of FPI and then another four $e$-foldings to cross over to FPII,
placing the energy scale at which one arrives near FPII at 300-700 TeV.
It is clear from Fig.~1 that these trajectories will have $a_0(m_Z)>-0.002$.
Numerical analysis shows that the locus of endpoints of such trajectories satisfies
\be
a_1(m_Z)=-0.00321 - 0.00052\, a_0(m_Z)\, .
\ee
For $a_0\approx 0.5$ this relation is still true within a few percent.

\begin{figure}
{\resizebox{0.8\columnwidth}{!}
{\includegraphics{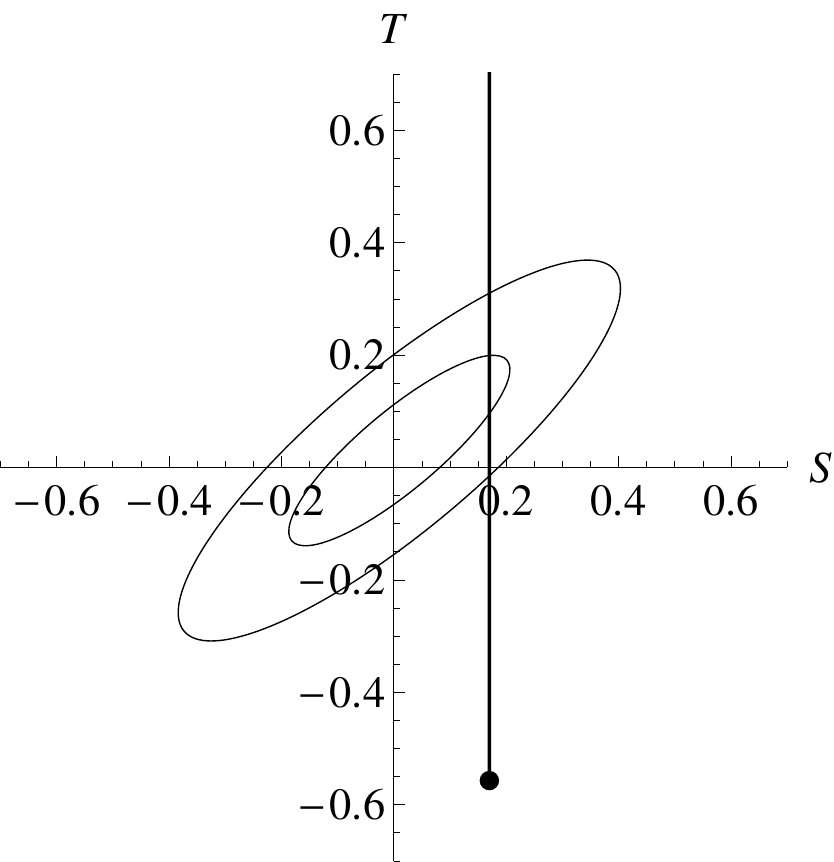}}
\caption{\small The half-line and the dot show the values permitted by asymptotic safety. 
The ellipses show the 1 and 2 $\sigma$ experimental bounds with $m_H$=117GeV \cite{pdg}.
}
}
\end{figure}

Using equations (\ref{sdef}) and (\ref{tdef}), this translates directly into a 
linear relation between $S$ and $T$, which is shown in Fig.~2,
and constitutes our main result.
The dot corresponds to the UV critical surface of FPI (\ref{point}), 
the half-line to the UV critical surface of FPII. 
Note that the condition of asymptotic safety essentially fixes $a_1$, and hence $S$, 
leaving $T$ arbitrary.

Renormalizable trajectories represent UV complete theories.
We see that within this model there are such trajectories 
that are in agreement with the experimental data: 
$S=0.01\pm0.10$ and $T=0.03\pm0.11$.
They pass near FPI at scales $\approx 10$ TeV and then veer towards FPII.
There, the custodial symmetry is strongly broken, as witnessed by the large
value $a_0\approx 0.5$.
This could be an important (and unexpected) clue about the UV behavior of the theory.
In this model the conformal (FP) behavior sets in at energies that are
probably too high to make a direct observation possible at LHC,
but there may be other signatures.
We will return to this and related questions elsewhere \cite{fptvz}.

We would like to thank O. Zanusso for discussions.



\begin{thebibliography}{99}
 
  
\bibitem{ab} 
T. Appelquist, C. Bernard, Phys. Rev. {\bf D22} 200 (1980);\\
A.C. Longhitano, Phys. Rev. {\bf D22} 1166 (1980).  

\bibitem{Weinberg}
 S.~Weinberg, ``Critical Phenomena For Field Theorists,'' \\
 Lectures presented at Int. School of Subnuclear Physics, Ettore Majorana, Erice, Sicily, Jul 23 - Aug 8, 1976.

\bibitem{reviews}
S. Weinberg, In {\it General Relativity: An Einstein centenary survey}, 
ed. S.~W. Hawking and W. Israel, pp.790--831, Cambridge University Press (1979);\\
M. Niedermaier and M. Reuter, Living Rev. Relativity 9, 5  (2006);\\
R. Percacci, 
in ``Approaches to Quantum Gravity: Towards a New Understanding of Space, Time and Matter'' 
ed. D. Oriti, Cambridge University Press (2009). 

\bibitem{Gies}
H. Gies and M.M. Scherer, 
Eur. Phys. J. {\bf C66} 387-402 (2010) 
arXiv:0901.2459 [hep-th];\\
H. Gies, S. Rechenberger and M.M. Scherer, 
Eur. Phys. J. {\bf C66} 403-418 (2010) 
arXiv:0907.0327 [hep-th];\\
X. Calmet, arXiv:1012.5529 [hep-ph].

\bibitem{codello}
  A.~Codello and R.~Percacci,
  Phys.\ Lett.\  B {\bf 672} 280 (2009)
  [arXiv:0810.0715 [hep-th]].
 
\bibitem{hasenfratz}
P. Hasenfratz, Nucl. Phys. {\bf B321} 139-162 (1989);\\
R.~Percacci and O.~Zanusso,
Phys. Rev. {\bf D81} 065012 (2010) [arXiv:0910.0851 [hep-th]].
  
\bibitem{fptz}
M. Fabbrichesi, R. Percacci, A. Tonero and O. Zanusso, Phys. Rev. {\bf D83} 025016 (2011).
 
\bibitem{peskin}
M.E. Peskin, T. Takeuchi,
Phys. Rev. Lett. {\bf 65} 964-967 (1990);
Phys. Rev. {\bf D46} 381-409 (1992). 
 
\bibitem{fptvz}
M. Fabbrichesi, R. Percacci, A. Tonero, L. Vecchi, O. Zanusso in preparation
 
\bibitem{feruglio}
F. Feruglio, 
2nd national seminar of theoretical physics, Parma, 1-12 september 1992, arXiv:hep-ph/930128.

\bibitem{herreroruiz}
M.J. Herrero, E. Ruiz-Morales, Nucl. Phys. {\bf B418} 431-455 (1994).

\bibitem{bagger}
J.A. Bagger, A.F. Falk, M. Swartz, 
Phys. Rev. Lett. {\bf 84} 1385-1388 (2000)
[arXiv: hep-ph/9908327].

\bibitem{pdg}
K. Nakamura et al. (Particle Data Group), J. Phys. G 37, 075021 (2010)


\end{thebibliography}
\end{document}